\documentclass[a4paper,12pt]{article}
\pdfoutput=1

\usepackage{amsmath}
\usepackage{amssymb}
\usepackage{mathrsfs}
\usepackage{bbm}
\usepackage{graphicx,subfigure}
\usepackage[numbers,sort&compress]{natbib}

\numberwithin{equation}{section}

\newlength{\dinwidth}
\newlength{\dinmargin}
\begin{document}

\title{\bf Constraints on anomalous $tcZ$
  coupling from $\bar B\to \bar K^*\mu^+\mu^-$ and $B_s \to
  \mu^+\mu^-$ decays}
\bigskip

\author{Hui Gong$^{1}$, Ya-Dong Yang$^{1,2}$ and Xing-Bo Yuan$^{1,2}$\\
{ $^1$\small Institute of Particle Physics, Central China Normal University, Wuhan, Hubei 430079, P.~R. China}\\
{ $^2$\small Key Laboratory of Quark \& Lepton Physics, Ministry of
  Education, Central China Normal University,}\\
{\small Wuhan, Hubei 430079, P.~R. China}}

\date{}
\maketitle
\bigskip\bigskip
\maketitle
\vspace{-1.5cm}

\begin{abstract}
{\noindent}
In this paper, we analyze the possible anomalous $tcZ$ coupling
effects in the $b \to s $ mediated decays $\bar B \to \bar K^* \mu^+ \mu^-$ and $B_s
\to \mu^+ \mu^-$. After exploiting
the available experimental data, the combined
constraints on the anomalous coupling $X_{ct}^L$ are derived. It is
found that, the bound on the magnitude $|X_{ct}^L|$ is dominated by the
branching ratios of these two decays. Furthermore, one sign-flipped solution is
excluded by the longitudinal fraction of $\bar B \to \bar K^* \mu^+
\mu^-$ at the low dilepton mass
region. After considering the combined constraints, for general complex coupling $X_{ct}^L$, the predicted upper bound on $\mathcal B (t \to
cZ)$ are compatible with that from the recent CMS direct search. In
particular, for
the case of real
coupling $X_{ct}^L$, the upper bound reads $\mathcal B (t \to cZ)<6.3\times
10^{-5}$, which is much lower than the current CMS bound but still
accessible at the LHC. With improved measurements at
the LHC, the colser correlations between the $t \to cZ$ and $b \to s$
mediated (semi-) leptonic decays are expected in the near future.
\end{abstract}

\newpage

\section{Introduction}
\label{sec:intro}
In the Standard Model (SM), the flavor-changing neutral current (FCNC)
transitions are forbidden at tree level and highly suppressed at
one-loop level due to the Glashow-Iliopoulos-Maiani (GIM)
mechanism \cite{GIM}. Such processes may receive competing contributions from
possible new physics (NP) beyond the SM, as a result of which the
expected rates related to these processes can be significantly altered. Thus the FCNC processes are promising probes of the SM and its extensions.

For the top quark in particular, the FCNC decays $t \to qZ$ (where $q$
denotes a $c$- or $u$-flavored quark) are
predicted to be far below the detectable level within the SM, with branching ratios
of order of $10^{-10}$ \cite{SM:ratio:1,SM:ratio:2}. However, there are various NP models
that may enhance these processes significantly \cite{top-quark
  physics:1,top-quark physics:2}. This makes any positive signal of these decays
an indirect evidence of NP beyond the SM. Search for the top quark FCNC decays has been performed at the
Tevatron \cite{exp:FCNC:CDF,exp:FCNC:D0} and the LHC \cite{exp:FCNC:ATLAS, exp:FCNC:CMS}. The best upper
limits on branching ratio of $\mathcal B (t\to qZ)<0.24\%$ at $95\%$
C.L. are recently
established by the CMS collaboration \cite{exp:FCNC:CMS}. Improved
direct searches will be available at the LHC due to its large top
sample in prospect. The discovery potential of $\mathcal B (t\to qZ)$ is of
the order $10^{-4}$ at the ATLAS \cite{discovery potential:ATLAS:1,
  discovery potential:ATLAS:2} and the CMS \cite{discovery potential:CMS}.

However, when studying the phenomena of the $t\to q Z$ transition, or
equivalently an effective anomalous $tqZ$ coupling, at high-energy colliders, the
low-energy processes involving the top quark loops should also be
taken into account \cite{top:FCNC:1,top:FCNC:2,top:FCNC:3,top:FCNC:4,top:FCNC:5,top:FCNC:6,top:FCNC:7,top:FCNC:8,top:FCNC:9}. In our previous works \cite{our work:1,our work:2, our work:3}, we have investigated the top quark anomalous
coupling effects in rare B and K-meson decays. For the anomalous $tcZ$
coupling in particular, it is found that the dominant constraints come from the $B_s\to
\mu^+\mu^-$ decay. As another $b \to s \mu^+ \mu^-$ process, the
$\bar B \to \bar K^* \mu^+\mu^-$ decay has been investigated in the literature \cite{Ali,QCDF:1,QCDF:2,TA:lowrecoil,charm-loop:1,charm-loop:2} and shown to be able to provide complementary
information about the potential NP contributions \cite{B2KVll:NP:1,B2KVll:NP:2,B2KVll:NP:3,B2KVll:NP:4,B2KVll:NP:5}. Its subsequent $K^* \to K \pi$
processes allow to offer a large number of observables in the fully
differential distribution through an angular analysis of the $K\pi \mu^+\mu^-$
final state \cite{angular:1,angular:2}. Furthermore, the hadronic
uncertainties in  some angular
observables cancel each other, which makes theoretical predictions precise
\cite{Virto:1,Virto:2}.
On the experimental side, the $\bar B \to \bar K^* \mu^+ \mu^-$ decays
have been measured by the experiments BaBar
\cite{exp:BaBar}, Belle \cite{exp:Belle}, CDF
\cite{exp:CDF,exp:CDF:ICHEP} and LHCb (with an integrated luminosity
of $0.37 \, {\rm fb}^{-1}$) \cite{exp:LHCb}. In the near
future the LHCb collaboration expects to improve these measurements by
using  an integrated luminosity of $1.5\, {\rm fb}^{-1}$ data \cite{exp:LHCb:ICHEP}.

Recently, the first evidence for the decay $B_s \to \mu^+ \mu^-$ has been announced
by the LHCb collaboration \cite{exp:Bs2ll}. The observed rate
$\mathcal B (B_s\to \mu^+\mu^-)=(3.2_{-1.2}^{+1.5})\times 10^{-9}$ is
in good agreement with the SM prediction. In this paper, motivated by
the LHCb result, we shall update the bound
on the anomalous $tcZ$ coupling obtained in our previous work \cite{our
  work:3} with this recent data. After performing a model-independent
study of the anomalous $tcZ$ coupling effects in the $\bar B \to \bar
K^* \mu^+ \mu^-$ decays, we derive the combined bounds on its strength
by these two decay modes. The
implications for the direct search of the rare $t\to cZ$ decays at the LHC are also discussed.

Our paper is organized as follows: In section 2, we introduce the
effective Lagrangian describing the anomalous
$tqZ$ interactions. Section 3 gives a short review on the $B_s \to
\mu^+ \mu^-$ and $\bar B \to \bar K^* \mu^+
\mu^-$ decays and also the anomalous $tcZ$ coupling effects in
these two decays. In section 4, we give our detailed numerical
results and discussions. We conclude in section 5. The relevant
formulae for our analysis are shown in the Appendices.

\section{Effective Lagrangian for anomalous $tqZ$ couplings}
From the viewpoint of effective field theory, the SM can be considered
as an effective low-energy theory of an underlying theory at a scale
$\Lambda$ which is much higher than the electroweak scale $v=246\,{\rm
  GeV}$ \cite{EFT}. The NP effects above the electroweak scale can be encoded in
the higher dimensional interaction terms involving only the SM fields
and invariant under the SM gauge symmetry. In particular, the FCNC transitions $t\to
qZ$ can be described by a few dimension 6 operators \cite{EFT:operator:1,EFT:operator:2}. These operators
can contribute to the $tqZ$ vertices, resulting an equivalent
description by the effective Lagrangian \cite{EFT:operator:1,EFT:Lagrangian:1,EFT:Lagrangian:2}
\begin{align}\label{eq:effective Lagrangian}
  \mathcal L_{tqZ}=&+\frac{g}{2\cos\theta_W}\bar q
  \gamma^\mu(X_{qt}^LP_L+X_{qt}^RP_R)tZ_\mu\nonumber\\
  &+\frac{g}{2\cos \theta_W}\bar q
  \frac{i\sigma^{\mu\nu}p_\nu}{m_Z}(\kappa_{qt}^LP_L+\kappa_{qt}^RP_R)tZ_\mu+{\rm
  h.c.},
\end{align}
with $P_{L,R}=(1\mp \gamma^5)/2$. The dimensionless couplings
$X_{qt}^{L,R}$ and $\kappa_{qt}^{L,R}$ are complex generally and can
be written in terms of its magnitude and phase as, for example,
$X_{ct}^L\equiv \lvert X_{ct}^{L} \rvert e^{i\theta_{ct}^{L}}$. This
Lagrangian has been employed in phenomenological analyses related to
top-quark physics \cite{top-quark physics:1,top-quark physics:2}.
\section{Theoretical Framework}
In this section, we shall first introduce the theoretical framework of
the $B_s \to \mu^+ \mu^-$ and $\bar B
\to \bar K^* l^+ l^-$ decays, and then discuss the anomalous $tqZ$
coupling effects in these two decays.
\subsection{Effective Hamiltonian}
In the SM, the effective Hamiltonian for the $b\to s l^+l^-$
transitions read \cite{Hamiltonian:1,Hamiltonian:2}
\begin{align}\label{eq:Hamiltonian}
  \mathcal H_{\rm eff}=-\frac{4G_F}{\sqrt 2}\left(\lambda_t\mathcal
  H_{\rm eff}^{(t)}+\lambda_u\mathcal H_{\rm eff}^{(u)}\right),
\end{align}
with the products of the CKM matrix elements $\lambda_q \equiv V_{qb}V_{qs}^*$ and
\begin{align}
  \mathcal H_{\rm eff}^{(t)}=C_1\mathcal O_1^c+C_2\mathcal
   O_2^c+\sum_{i=3}^{10}C_i\mathcal O_i,\qquad   \mathcal H_{\rm eff}^{(u)}=C_1(\mathcal O_1^c-\mathcal
  O_2^u)+C_2(\mathcal O_2^c-\mathcal O_2^u),
\end{align}
where the explicit expressions of $\mathcal O_{1-6}$ can be found in
ref.~\cite{Hamiltonian:1,Hamiltonian:2}. The
Wilson coefficients, which contain the short-distance physics, can be
calculated perturbatively at the high scale $\mu=\mu_W$. At the low scale
$\mu=\mu_b$, their values are obtained by means of QCD renormalization
group equations , which has
been performed at NNLL accuracy \cite{Hamiltonian:2, ADM:1,ADM:2,ADM:3}. For
the $B_s \to \mu^+\mu^-$ and $\bar B\to \bar K^* l^+ l^-$ decays, the
electromagnetic dipole operator and semileptonic four-fermion
operators are more relevant \cite{Ali}
\begin{align} \label{eq:operator}
  \mathcal O_7=\frac{e}{g_s^2}m_b(\bar s \sigma_{\mu\nu} P_R b)
  F^{\mu\nu},\quad  \mathcal O_9=\frac{e^2}{g_s^2}(\bar s \gamma_\mu P_L
  b) (\bar l
  \gamma^\mu l),\quad
  \mathcal O_{10}=\frac{e^2}{g_s^2}(\bar s \gamma_\mu P_L b) (\bar l
  \gamma^\mu \gamma_5 l),
\end{align}
where $g_s$ is the strong coupling constant. The remaining operators enter the matrix elements at higher
orders through the following combinations, named effective Wilson
coefficients \cite {effective Wilson coefficient},
\begin{align}\label{eq:effective WCs}
  C_7^{\rm
    eff}&=\frac{4\pi}{\alpha_s}C_7-\frac{1}{3}C_3-\frac{4}{9}C_4-\frac{20}{3}C_5-\frac{80}{9}C_6,\nonumber\\
  C_9^{\rm eff}&=\frac{4\pi}{\alpha_s}C_9+Y(q^2),\nonumber\\
  C_{10}^{\rm eff}&=\frac{4\pi}{\alpha_s}C_{10},
\end{align}
where the function $Y(q^2)$ is defined in eq.~(\ref{eq:Y}).
\subsection{Theoretical formalism}
\subsubsection{$B_s \to \mu ^+ \mu^-$}
The rare decay $B_s \to \mu^+ \mu^-$ is one of the most powerful
probes in the search for deviations from the SM. In the effective
Hamiltonian of eq.~(\ref{eq:effective WCs}), only the operator $\mathcal
O_{10}$ is relevant to this process and the branching ratio is given
by \cite{Bs2ll:BR:1,Bs2ll:BR:2,Bs2ll:BR:3}
\begin{align}
\mathcal B (B_s\to\mu^+\mu^-)=\frac{G_F^2 \alpha_e^2 }{64\pi^3} m_{B_s}^3 f_{B_s}^2
\lvert \lambda_t \rvert ^2   \tau_{B_s}\sqrt{1-\frac{4
    m_\mu^2}{m_{B_s}^2}} \cdot \frac{4 m_\mu^2}{m_{B_s}^2} \cdot \left\lvert
C_{10}^{\rm eff} \right\rvert^2,
\end{align}
where $f_{B_s}$ is the $B_s$ meson decay constant. Recently,  a sizable width difference
$\Delta\Gamma_s$ between the $B_s$ mass eigenstates has been
measured at the LHCb \cite{ys}
\begin{align}
    y_s\equiv \frac{\Gamma_s^{\rm L}-\Gamma_s^{\rm H}}{\Gamma_s^{\rm
        L}+\Gamma_s^{\rm H}}=\frac{\Delta\Gamma_s}{2\Gamma_s}=0.088\pm0.014,
\end{align}
where $\Gamma_s$ is the inverse of the $B_s$ mean lifetime
$\tau_{B_s}$. Interestingly, it has been pointed out in
ref.~\cite{Bs2ll:BR:4:1,Bs2ll:BR:4:2} that the sizable width difference should be
taken into account in the evaluation of the branching ratio to compare
with its experimential measurement, thus the measured branching ratio
should be the time-integrated one which reads
\begin{align}\label{eq:correction factor}
  \overline{\mathcal B} (B_s \to \mu^+\mu^-)=
  \left[\frac{1+\mathcal  A_{\Delta\Gamma}y_s}{1-y_s^2}\right] \mathcal B (B_s \to\mu^+\mu^-),
\end{align}
where the
observable $\mathcal A_{\Delta\Gamma}$, equals to +1 in the SM, may vary in the interval
$[-1,+1]$ in the presence of NP and can be extracted from
time-dependent measurement of $B_s\to \mu^+\mu^-$.
\subsubsection{$\bar B \to \bar K^* l^+ l^-$}\label{sec:B2KVll}
The $\bar B \to \bar K^* l^+ l^-$ decays are induced by the $b \to s l^+ l^-$
transition at quark level and provide constraints on the semileptonic
operators $\mathcal O_{9,10}$. In this paper, we focus on the differential branching ratio,
forward-backward asymmetry and longitudinal fraction at both the large and
 the low hadronic recoil, which can be
built from transversity amplitudes
\begin{align}\label{eq:observables}
 \frac{ {\rm d}\Gamma }{{\rm d}q^2}&=|A_0^L|^2+|A_\perp^L|^2+|A_\parallel^L|^2+(L\leftrightarrow R),\nonumber\\
A_{\rm FB}&=\frac{3\beta_l}{2}\frac{{\rm Re}(A_\parallel^LA_\perp^{L*})-{\rm Re}(A_\parallel^RA_\perp^{R*})}{{\rm d}\Gamma/{\rm d}q^2},\nonumber\\
F_{\rm L}&=\frac{|A_0^L|^2+|A_0^R|^2}{{\rm d}\Gamma/{\rm d}q^2},
\end{align}
where $q^2$ is the dilepton invariant mass squared and
$\beta_l=\sqrt{1-4m_l^2/q^2}$ is the phase space factor. After the first treatment by naive
factorization \cite{Ali}, it has been shown that a systematic theoretical
description using QCD factorization (QCDF) apples in the large
hadronic recoil region $1\,{\rm GeV}^2 \lesssim q^2 \ll 4m_c^2\approx 7\,{\rm GeV}^2$
\cite{QCDF:1,QCDF:2}. For the low recoil region $q^2 \gtrsim 15\,{\rm GeV}^2$, an approach based on an
operator product expansion in $1/m_b$ and in $1/\sqrt{q^2}$ with
improved Isgur-Wise form factor relations has been developed
\cite{Buchalla:1998,Buchalla:2011,Grinstein:2002,Grinstein:2004,TA:lowrecoil}. These
theoretical treatments and the corresponding expressions of the
transversity amplitudes $A_{\perp,\parallel,0}^{L,R}$ are shown in
appendix B. It is noted that the observables ${\rm d}\Gamma/{\rm d}q^2$ and
$A_{\rm FB}$ at low recoil only depend on the two combinations of
Wilson coefficients $\rho_1$ and $\rho_2$ defined in eq.~(\ref{eq:rho}), while the $F_{\rm L}$ are independent of Wilson coefficients.

For the $7\,{\rm GeV}^2 \lesssim q^2 \lesssim 15\, {\rm GeV}^2$ region, the
large quark-hadron duality violations caused by narrow $c \bar c$
resonances \cite{quark-hadron duality} and
the hadronic backgrounds from $B \to K^* \psi^{(\prime)}$ make it
difficult to give reliable theoretical predictions from the first
principle. In the $q^2 \lesssim 1 \,{\rm GeV^2}$ region, the factorization formulae suffer from end-point divergences (at $q^2 \approx \Lambda_{\rm QCD}^2$) \cite{QCDF:1,QCDF:2} and there could also be
(unknown) resonance contributions from $\rho$ or other mesons as
discussed in ref.~\cite{Buras}. Additionally, theoretical predictions
dependent largely on the Wilson coefficient $C_7^{\rm eff}$, which is
bounded more stringently by the $b\to s \gamma$ processes. Therefore, we do not include these two regions in
our analysis.

\subsection{Anomalous $tqZ$ coupling effects}
The effective $tqZ$ vertices in eq.~(\ref{eq:effective Lagrangian}) affect the $b \to s l^+ l^- $ processes
through entering the $bsZ$-penguin diagram and result in the following
effective Hamiltonian \cite{our work:3}
\begin{align}
 \mathcal H_{\rm eff}^{\rm NP}=-\frac{4G_F}{\sqrt 2} \lambda_t
 \frac{\alpha_e}{2\pi\sin^2\theta_W} C^{\rm NP} (\bar s \gamma_\mu P_L b)(\bar l \gamma^\mu P_L l),
\end{align}
or in the operator basis in eq.~(\ref{eq:operator})
\begin{align}
  \mathcal H_{\rm eff}^{\rm NP}=-\frac{4G_F}{\sqrt 2} \lambda_t
  \frac{\alpha_s}{4\pi}\frac{C^{\rm NP}}{\sin^2\theta_W}(\mathcal O_9 -
  \mathcal O_{10}).
\end{align}
The matching coefficient $C^{\rm NP}$ has been calculated in the unitary gauge
with the modified minimal subtraction ($\overline{\rm MS}$) scheme
\cite{our work:3}. It is found that, besides the left-handed vector
current $\bar c \gamma^\mu P_L t Z_\mu$ all the contributions from the
$tqZ$ anomalous interactions in the effective Lagrangian
eq.~(\ref{eq:effective Lagrangian}) can be safely neglected, then the coefficient reads
\begin{align}
  C^{\rm NP}=&-\frac{1}{8} \frac{V_{cs}^*}{V_{ts}^*}\biggl[\left(-x_t \ln
    \frac{m_W^2}{\mu^2}+\frac{3}{2}+x_t-x_t \ln
    x_t\right) X_{ct}^L+ \mathcal O \left( \frac{m_c}{m_W} \right) X_{ct}^R \biggr],
\end{align}
with $x_t\equiv\bar m_t^2 / m_W^2$. Compared with the SM case, the
coefficient $C^{\rm NP}$ is
enhanced by a large
CKM factor $V_{cs}^*/ V_{ts}^*$, which makes the $b \to s l^+ l^-$
transitions sensitive to the anomalous $tcZ$ coupling.

Normalized to
the effective Hamiltonian eq.~(\ref{eq:Hamiltonian}), the anomalous
$tcZ$ coupling effects result in the following deviations
\begin{align}
  C_9^{\rm eff}&\to \widetilde{C}_9^{\rm eff}=C_9^{\rm eff}+ \frac{C^{\rm NP}}{\sin^2
    \theta_W}=+4.29\left(1+17.3|X_{ct}^L|e^{i(\theta_{ct}^L+\beta_s)}\right)+Y(q^2),\nonumber\\
  C_{10}^{\rm eff}
  &\to \widetilde{C}_{10}^{\rm eff}=C_{10}^{\rm eff} -\frac{C^{\rm NP}}{\sin^2 \theta_W}=-4.22\left(1+17.6|X_{ct}^L|e^{i(\theta_{ct}^L+\beta_s)}\right),
\end{align}
where the SM Wilson coefficients are the NNLL numerical values and the
phase $\beta_s \approx 1.04^\circ$ can be obtained from its definition
$\beta_s\equiv-\arg (-V_{cs}V_{cb}^*/V_{ts}V_{tb}^*)$. It is noted
that, in the transversity amplitudes $A_{\perp,\parallel,0}^R$, the
anomalous coupling only enters the combination $(C_9^{\rm
  eff}+C_{10}^{\rm eff})$ and their effects would be tiny.

For the $B_s\to \mu^+\mu^-$ decay in particular, the deviation of the
Wilson coefficient $C_{10}^{\rm eff}$ also enters the observable $\mathcal A_{\Delta\Gamma}$ \cite{Bs2ll:BR:4:1,Bs2ll:BR:4:2},
\begin{align}
  \mathcal A_{\Delta\Gamma}=\cos\left[2\arg\left(\frac{\widetilde C_{10}^{\rm eff}}{C_{10}^{\rm eff}}\right)\right]=\cos\left[2\arg\left(1+17.6|X_{ct}^L|e^{i(\theta_{ct}^L+\beta_s)}\right)\right].
\end{align}
This NP effects would result in a tiny suppression (up to 2\%) on the branching ratio.

For the rare $t \to c Z$ decay mediated by
anomalous $tcZ$ coupling, the branching ratio has been calculated at NLO \cite{t2qZ:LO,t2qZ:NLO:1,t2qZ:NLO:2,t2qZ:NLO:3,t2qZ:NLO:4}. We shall follow the
treatment of ref.~\cite{our work:3} and only consider the LO results.
\section{Numerical results and discussions}
With the theoretical framework discussed in previous sections and the
input parameters collected in table \ref{tab:input}, we shall present our
numerical results and discussions in this section.

\subsection{The SM predictions} \label{sec:SM predictions}

\begin{table}[t]
  \centering
  \begin{tabular}{|l l l|l l l|}
    \hline
    $G_F$&$1.1663787\times 10^{-5}\,{\rm GeV}^{-2}$&\cite{input:PDG} &$m_W$&$ (80.385\pm 0.015) \,{\rm GeV}$&\cite{input:PDG}\\
    $ \sin^2\theta_W$&$0.23146$&\cite{input:PDG}&$m_Z$&$(91.1876\pm 0.0021)\,{\rm GeV}$&\cite{input:PDG}\\
    $\alpha_s(m_Z)$&$0.1184\pm 0.0007$ & \cite{input:PDG}& $m_t^{\rm
      pole}$&$(173.18\pm 0.94)\,{\rm GeV}$&\cite{input:top}\\
   $\alpha_e(m_Z)$& 1/127.944 &\cite{input:PDG}&$m_c(m_c)$&$(1.275\pm
    0.025)\,{\rm GeV}$&\cite{input:PDG}\\
   $\alpha_e(m_b)$&$1/133$& \cite{TA:lowrecoil}     & $m_b(m_b)$&$(4.18\pm 0.03)\,{\rm
    GeV}$&\cite{input:PDG}\\
    $A$&$0.812_{-0.022}^{+0.015}$&\cite{input:CKM:CKMfitter}&$m_{K^*}$&$895.94\, {\rm
      MeV}$&\cite{input:PDG}\\
    $\lambda$&$0.22543_{-0.00095}^{+0.00059}$&\cite{input:CKM:CKMfitter}&$m_{B^0}$&$5279.58\, {\rm MeV}
    $&\cite{input:PDG}\\
    $  \bar\rho$&$0.145 \pm
    0.027$&\cite{input:CKM:CKMfitter}&$m_{B_s}$&$5366.77\, {\rm MeV}$&\cite{input:PDG}\\
     $\bar\eta$&$0.343 \pm 0.015$&\cite{input:CKM:CKMfitter}&$
     \tau_{B^0}$&$(1.519\pm 0.007)\, {\rm ps}$&\cite{input:PDG}\\ \cline{1-3}
     $f_{B_s}$&$(227.6\pm 5.0)\,{\rm MeV}$
     &\cite{input:fBs}&$\tau_{B_s}$&$(1.466\pm 0.031)\, {\rm ps}$&\cite{input:PDG}\\\hline
     \multicolumn{5}{|c}{$\overline{\mathcal B} (B_s \to \mu^+
       \mu^-)=(3.2_{-1.2}^{+1.5})\times 10^{-9}$\,\, ($\in
       [1.3,5.8] \times 10^{-9}$ at 90\% C.L.) }& \cite{exp:Bs2ll}\\
    \hline
  \end{tabular}
  \caption{The relevant input parameters and experimental data used in our numerical
    analysis.}
  \label{tab:input}
\end{table}

\begin{figure}[t]
  \centering
  \subfigure{\includegraphics[width=0.7\textwidth]{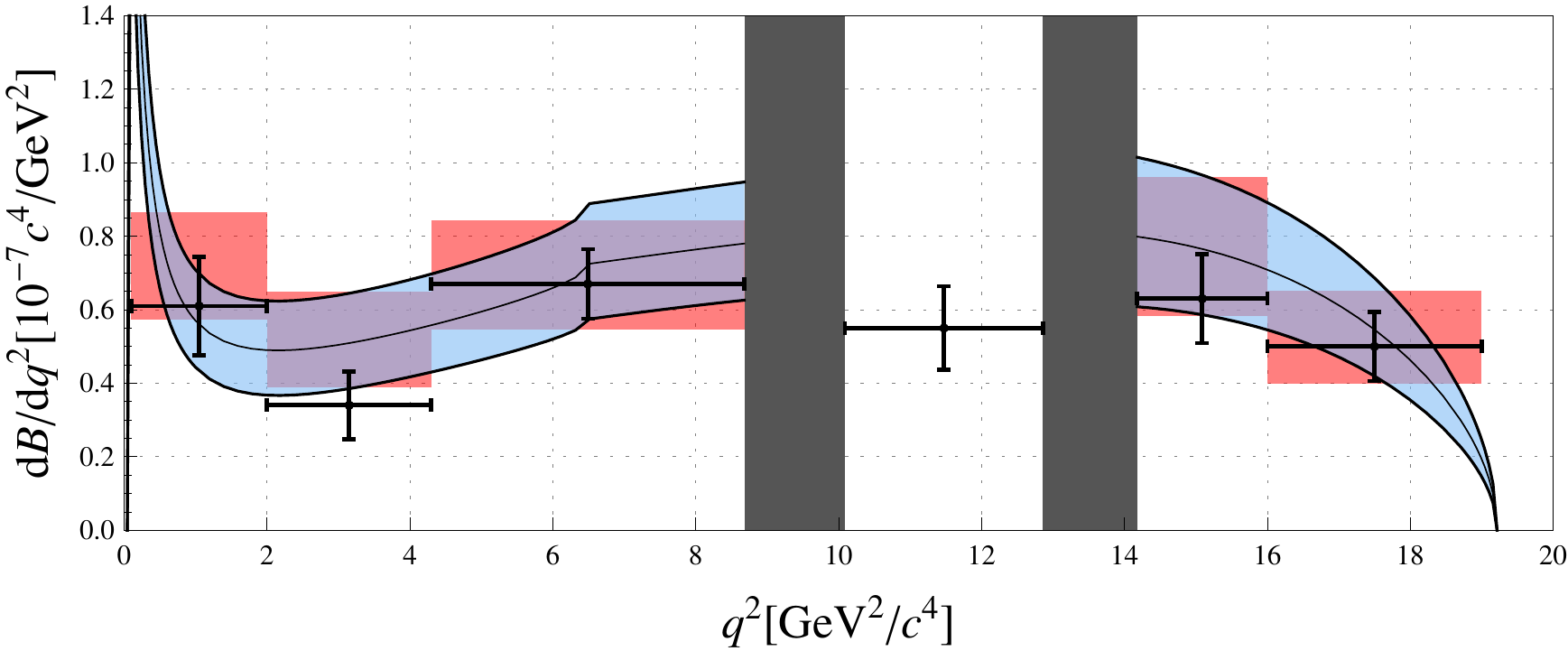}}
  \subfigure{\includegraphics[width=0.7\textwidth]{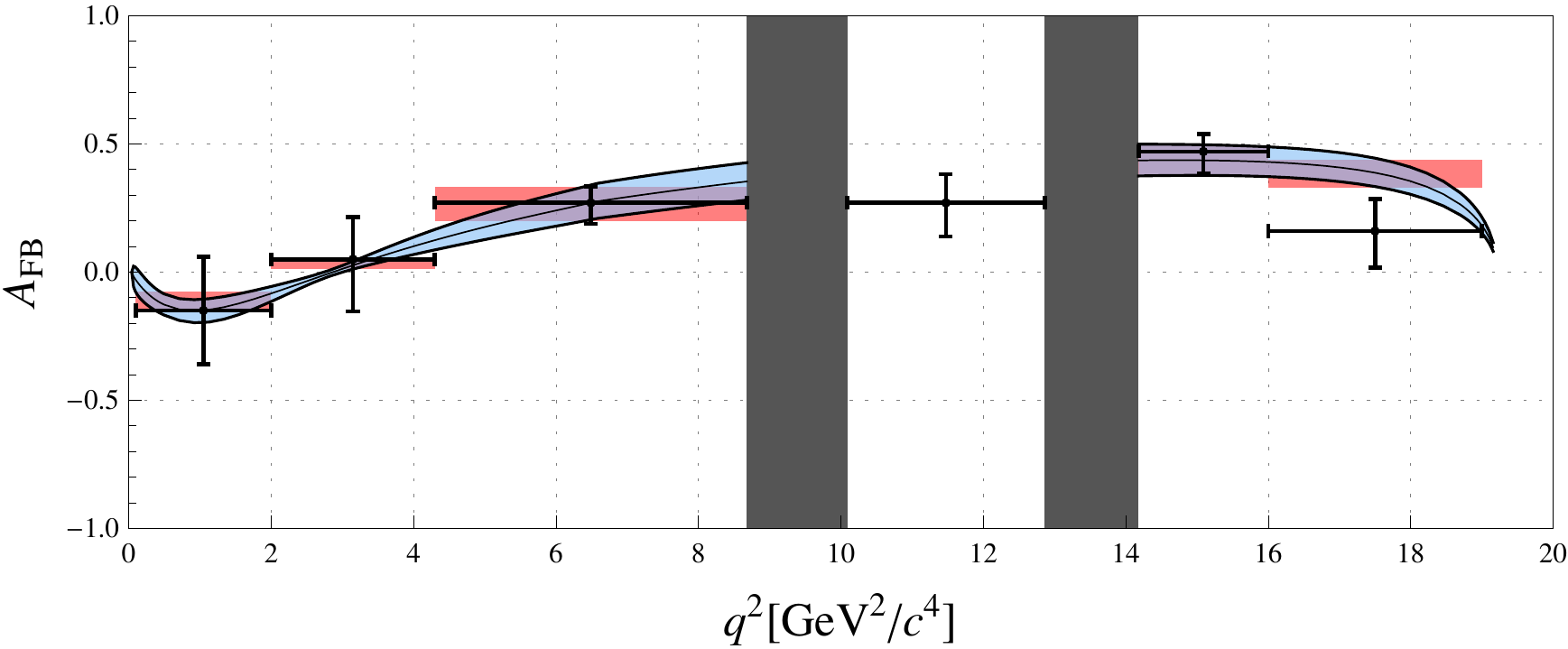}}
  \subfigure{\includegraphics[width=0.7\textwidth]{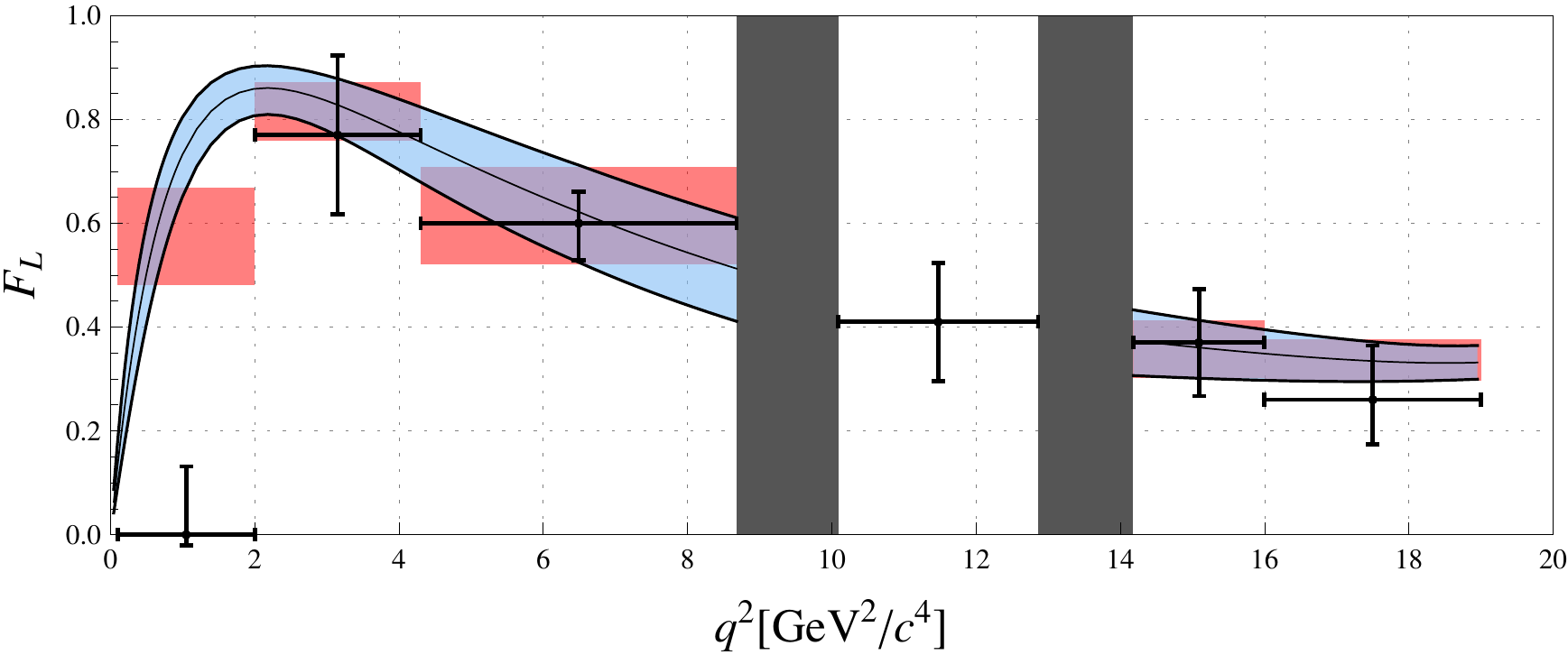}}
  \caption{The SM predictions for the  ${\rm d}\mathcal B /
  {\rm d} q^2$, $A_{\rm FB}$ and $F_{\rm L}$ with the theoretical
  uncertainties (the bands) versus the experimental measurements by
  LHCb \cite{exp:LHCb}. The rate-averaged observables are indicated by
  the red regions.}
  \label{fig:SM}
\end{figure}
For the decay $\bar B\to \bar K^* \mu^+\mu^-$, our SM predictions for the branching ratio ${\rm
  d}\mathcal B /{\rm d} q^2$, the forward-backward asymmetry $A_{\rm
  FB}$ and the longitudinal polarization fraction $F_{\rm L}$ with the
available data from LHCb \cite{exp:LHCb} are shown in
figure \ref{fig:SM}. For the integrated observables, we take the ratio in
eq.~(\ref{eq:observables}) after integrating its numerator and denominator
separately. This definition agrees with the one used in the experimental
measurements \cite{TA:lowrecoil}.

For the theoretical uncertainties, we follow
closely the
treatment of ref.~\cite{TA:lowrecoil}. At large recoil, we employ
the naive factorization approach, therefore a real scale
factor varying within $\pm 10 \%$ is added to each of the transversity amplitudes
$A_{\perp,\parallel,0}^{L,R}$ in eq.~(\ref{eq:TA:largerecoil}) to account for the $\mathcal O (\alpha_s)$ and $\mathcal O
(\Lambda/m_b)$ subleading QCDF corrections. Similarly, at low recoil, the
$\mathcal O (\alpha_s\Lambda/m_b)$
subleading corrections to each of the transversity amplitudes
$A_{\perp,\parallel,0}^{L,R}$ in eq.~(\ref{eq:TA:lowrecoil}) are estimated by real scale factors
varying within $\pm 5  \%$. For the uncertainties due to the $\mathcal O
(\Lambda/m_b)$ subleading corrections to the improved Isgur-Wise
relations and the $\mathcal O(m_{K^*}/m_B)$ neglected kinematical
factors, three real scale factors with $\pm 20\%$ uncertainties are assigned to the $\kappa C_7^{\rm
  eff}$ term for $A_{\perp,\parallel,0}^{L,R}$ in
eq.~(\ref{eq:TA:lowrecoil}), respectively. We obtain the theoretical uncertainties
by varying each of the input parameters and the real scale factors mentioned
above within its respective range and adding the individual
uncertainty in quadrature.

From figure \ref{fig:SM}, it can be seen that the theoretical
uncertainties for the branching ratio are about 30\%, which mainly arise from
the $B \to K^*$ form factors.  However, in the angular observables forward-backward asymmetry and
longitudinal fraction, these hadronic
uncertainties cancel each other, which result in the relatively precise theoretical predictions.

With the theoretical uncertainties taken into account, the LHCb data
are well consistent with the SM predictions except the $F_{\rm L}$ in the
lowest-$q^2$ bin $q^2 \in [0.10,2.00]\,{\rm GeV}^2$. For this bin, it is noted that the
recent LHCb preliminary result of the $F_{\rm L}$ (with $\mathcal L=1\,{\rm fb}^{-1}$) \cite{exp:LHCb:ICHEP} deviates from their previous
result \cite{exp:LHCb} by about
$2.5\sigma$ and is in agreement with
the SM prediction.

For the $B_s \to \mu^+ \mu^-$ decay, with the up-to-date input
parameters listed in table \ref{tab:input}, we obtain the SM prediction
\begin{align}
  \overline{\mathcal B} (B_s\to \mu^+ \mu^-)= (3.71\pm 0.24)\times 10^{-9},
\end{align}
where the theoretical uncertainty is dominated by the decay constant
$f_{B_s}$ and the CKM matrix elements. As pointed out in
ref.~\cite{Bs2ll:BR:4:1,Bs2ll:BR:4:2}, the correction in
eq.~(\ref{eq:correction factor})  gives a 10\% enhancement of the branching ratio
(compared with the one without the effects of $B_s^0-\bar
B_s^0$ mixing). This value is in good agreement
with the observed rate at the LHCb, which will put severe constraints on
the anomalous $tcZ$ coupling as seen in the following analysis.
\subsection{Constraining anomalous $tcZ$ coupling}
In order to constrain the anomalous $tcZ$ coupling, we consider the SM
predictions with $2\sigma$ error bars and the experimental data with
$1\sigma$ error bar. For the $\bar B \to \bar K^* \mu^+ \mu^-$ decays,
as discussed in subsection \ref{sec:B2KVll} and \ref{sec:SM
  predictions}, the LHCb data \cite{exp:LHCb} in the following
four bins $q^2 \in [2.00,4.30]\,{\rm GeV}^2$, $[4.30,8.68]\,{\rm GeV}^2$,
$[14.18,16.00]\,{\rm GeV}^2$
and $[16.00,19.00]\,{\rm GeV}^2$ are included in our analysis, which
are also depicted in figure \ref{fig:SM}. For the decay $B_s \to
\mu^+ \mu^- $, we consider the recent LHCb measurements \cite{exp:Bs2ll} listed in
table \ref{tab:input}.

\begin{figure}[t]
  \centering
  \subfigure{\includegraphics[width=0.48\textwidth]{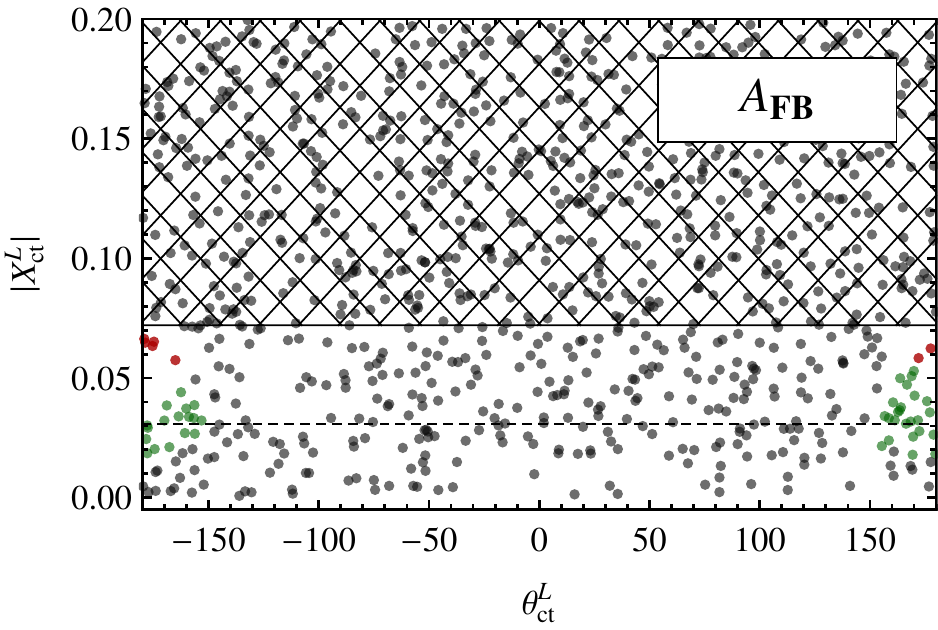}}\quad
  \subfigure{\includegraphics[width=0.48\textwidth]{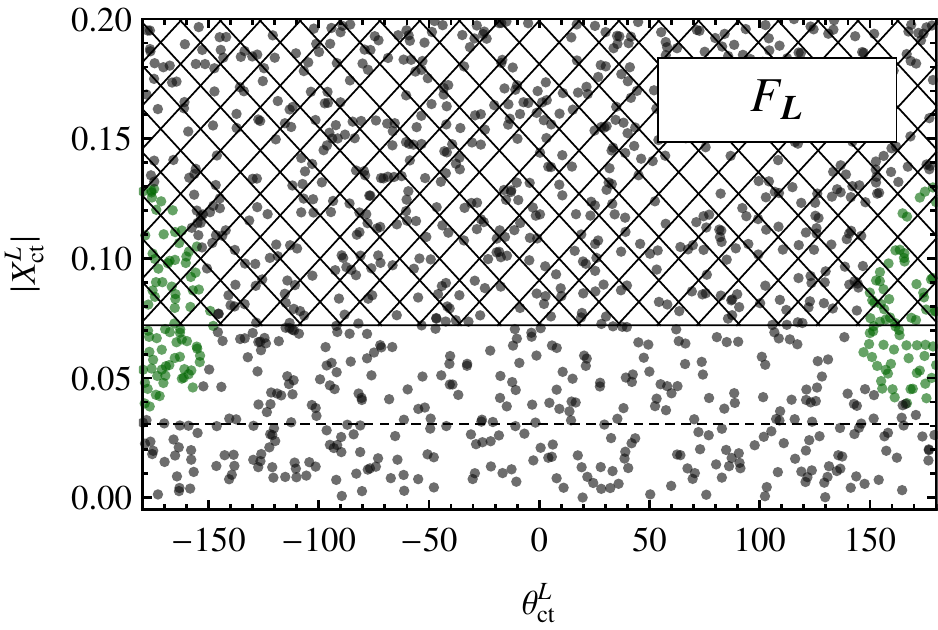}}
  \subfigure{\includegraphics[width=0.48\textwidth]{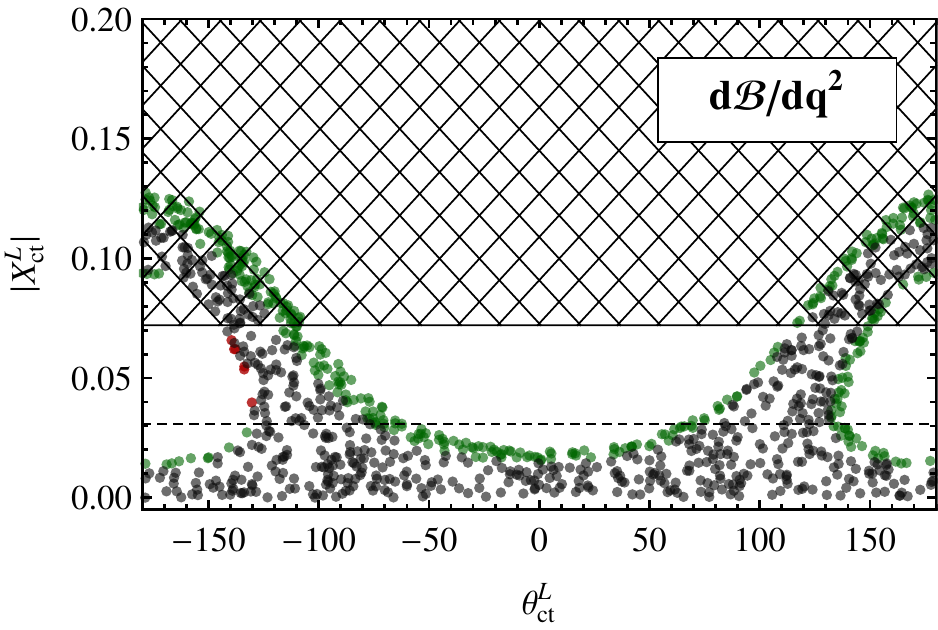}}\quad
  \subfigure{\includegraphics[width=0.48\textwidth]{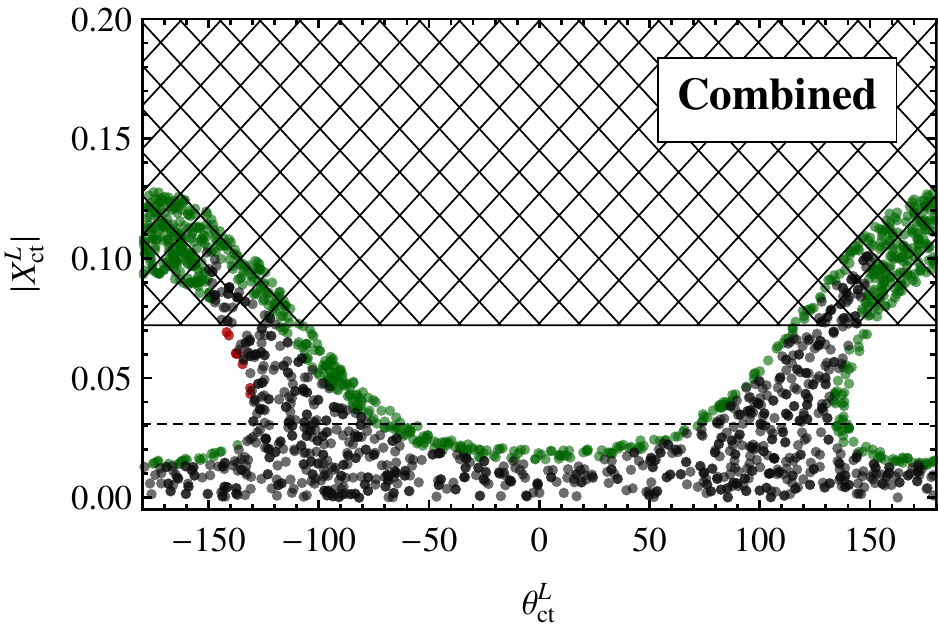}}
  \caption{The allowed regions of the anomalous coupling $|X_{ct}^L|$ as a
    function of $\theta_{ct}^L$ under the constraints from $\bar B
    \to \bar K^* \mu^+ \mu^-$ distributions ${\rm
      d}\mathcal B / {\rm d} q^2$, $A_{\rm FB}$ and $F_{\rm L}$ and their combinations by using the
    LHCb data \cite{exp:LHCb}. The allowed
    regions by the experimental data at large recoil (low recoil) are
    shown in red and
    black (green and black) points. Furthermore, the black points are
    allowed by all the data. The cross-hatched region are
    excluded by the CMS bound on $\mathcal B (t \to cZ)$ \cite{exp:FCNC:CMS}. The ATLAS $5 \sigma$ discovery potential
    at $\mathcal L=10\; {\rm fb}^{-1}$ \cite{discovery potential:ATLAS:1,
      discovery potential:ATLAS:2} is indicated by the dashed line. }
  \label{fig:NP}
\end{figure}

\begin{figure}[t]
  \centering
  \includegraphics[width=0.57\textwidth]{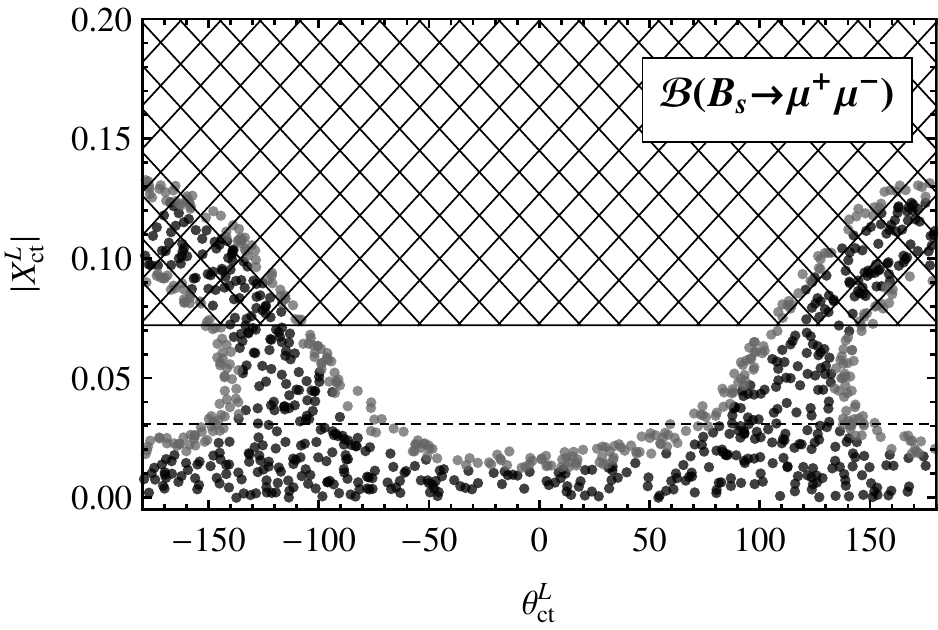}
  \caption{The upper bounds on the anomalous coupling $|X_{ct}^L|$ as a
  function of $\theta_{ct}^L$, constrained by the $\overline{\mathcal B} (B_s
  \to \mu^+\mu^-)$. The black (gray and black) region are the allowed
  parameter space obtained with
  $1\sigma$ experimental error bar (the experimental true
  value interval at $90\%$ C.L.). The other captions are
  the same as in figure \ref{fig:NP}.}
  \label{NP:Bs2ll}
\end{figure}

\begin{figure}[t]
  \centering
  \includegraphics[width=0.57\textwidth]{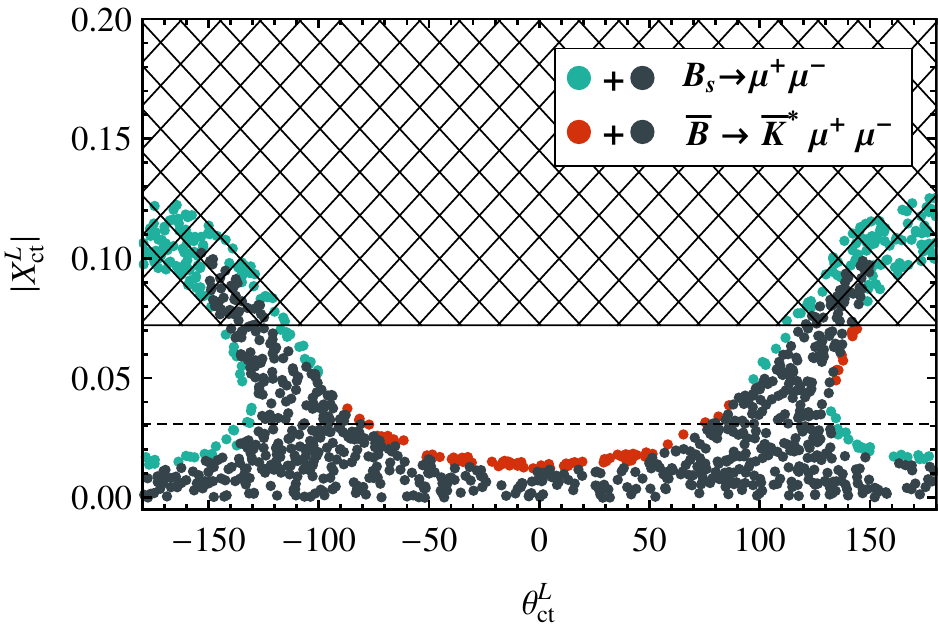}
  \caption{The combined upper bounds on the anomalous coupling
    $|X_{ct}^L|$ as a function of $\theta_{ct}^L$. The allowed regions
    by the experimental data of $B_s \to \mu^+ \mu^-$ ($\bar B \to
    \bar K^* \mu^+ \mu^-$) are indicated by both green and black (red
    and black) points. Furthermore, the black points are allowed by
    all the data. The other captions
    are the same as in figure \ref{fig:NP}.}
  \label{fig:final}
\end{figure}

The constraints on the anomalous coupling in the
$|X_{ct}^L|-\theta_{ct}^L$ plane by the ${\rm d}\mathcal B/
{\rm d}q^2$, $A_{\rm FB}$, $F_{\rm L}$ and their combinations in the $\bar B \to
\bar K^* \mu^+\mu^-$ decays are shown in figure \ref{fig:NP}. It can be seen that, the potentially large
$tcZ$ coupling effects are reflected in the stringent bound on its
magnitude $|X_{ct}^L|$, which is currently dominated by the differential
branching ratio. For a given value
$|X_{ct}^L|$,  the NP contribution is constructive to the SM one in the
region $\theta_{ct}^L\approx 0^\circ$, whereas their interference
becomes destructive in the region $\theta_{ct}^L\approx
\pm 180^\circ$. Therefore, there are two favored solutions under
the constraints of differential branching ratio in the $\theta_{ct}^L\approx \pm180^\circ$
region. The solution with larger $|X_{ct}^L|$ corresponds to the case
that the sign of
$(C_9^{\rm eff},C_{10}^{\rm eff})$ has been flipped by the anomalous
coupling. This solution is also not consistent with the CMS
bound as depicted in figure \ref{fig:NP}. For the
forward-backward asymmetry and longitudinal fraction,
they can not provide strong constraints on the anomalous
coupling. However, the longitudinal fraction at large recoil excludes
this sign-flipped solution. Exploiting the
$\bar B \to \bar K^* \mu^+ \mu^-$ data, the constraints from the
large recoil region are slightly more stringent than the
ones from low recoil and comparable to the latter.

For the
constraints on the anomalous $tcZ$ coupling by the $B_s \to \mu^+
\mu^-$ decays, after considering the recent LHCb data \cite{exp:Bs2ll}, we update our previous results \cite{our work:3} at figure \ref{NP:Bs2ll}. It can be seen that, the interference structure between
the SM and NP contributions manifested in the branching ratio is
similar to the one in the $\bar B \to \bar K^* \mu^+ \mu^-$
decays. Compared with our previous constraints, since the experimental
data of $B_s \to \mu^+ \mu^-$ are now double-bounded,
it excludes parts of the parameter space in the destructive region
$\theta_{ct}^L \approx \pm 180^\circ$ and leaves two solutions for $\lvert X_{ct}^L \rvert$.

The combined constraints obtained after considering all the
experimental data of both $B_s \to \mu^+ \mu^-$ and $\bar B \to \bar
K^* \mu^+ \mu^-$ decays are shown in figure \ref{fig:final}. In the parameter space of ($\theta_{ct}^L$, $\lvert X_{ct}^L\rvert$), benefited
from the angular observables, the $\bar B \to \bar K^* \mu^+
\mu^-$ process excludes the regions corresponding to
the sign-flipped solution. The constraints on the other part of the parameter space
provided by these two decays are almost the same stringent. The detailed
numerical results are listed in
table \ref{tab:result}. For general complex coupling $X_{ct}^L$, one can see that the predicted upper bound
$\mathcal B (t \to cZ)<4.85 \times 10^{-3}$ is compatible with the CMS
direct bound $\mathcal B (t \to cZ) <2.4 \times 10^{-3}$. For real coupling, the
corresponding bound $\mathcal B (t \to cZ)<6.3\times 10^{-5}$ is
below the CMS bound, but still of the same order as the $5\sigma$
discovery potential $\mathcal B (t \to cZ) \approx 4.4\times 10^{-4}$ of the LHC with an integrated luminosity of $10\, {\rm fb}^{-1}$.

\begin{table}[t]
  \centering
  \begin{tabular}{| c |c c c |c |}
    \hline
    &\multicolumn{3}{c|}{$\lvert X_{ct}^L \rvert$ constrained from
    }&our bound\\ \cline{2-4}
    $\theta_{ct}^L$&$B_s\to \mu^+ \mu^-$&$\bar B \to \bar K^* \mu^+ \mu^-$&CMS& $\mathcal B (t \to cZ)$\\ \hline
    $0^\circ$&$<0.012$&$<0.015$&$<0.072$&$<6.27\times 10^{-5}$\\
    $180^\circ$&$<0.125$&$<0.013$&$<0.072$&$<0.75\times 10^{-4}$\\
    $general$&$<0.125$&$<0.102$&$<0.072$& $<4.85\times 10^{-3}$\\
    \hline
  \end{tabular}
  \caption{The upper bounds on the magnitude $\lvert X_{ct}^L
    \rvert$ from the
    $\bar B \to \bar K^* \mu^+ \mu^-$ and $B_s \to \mu^+ \mu^-$ decays
    for some particular  phases
    $\theta_{ct}^L$. The corresponding predicted bounds on
    $\mathcal B (t \to c Z)$ are also given. The bounds in the fourth
    column are obtained from the direct search of the $t \to cZ$ decay
  at the CMS \cite{exp:FCNC:CMS}.}
  \label{tab:result}
\end{table}

\section{Conclusions}
In this paper, we have studied the effects of anomalous $tcZ$
coupling in the $\bar B \to \bar K^* \mu^+ \mu^-$ decays at both the
large and the low hadronic recoil region. With the recent LHCb measurements
of $B_s \to \mu^+ \mu^-$, the combined constraints on the
anomalous coupling $X_{ct}^L$ from these two
decays are derived. For general complex coupling, it is found that, the predicted upper limit of $\mathcal B(t\to cZ)$
is compatible with the CMS direct search. In particular, for real
coupling, the corresponding limit is below the current CMS bound, but
still stays in the accessible level of the LHC with an integrated
luminosity of $10\;{\rm fb}^{-1}$.

It has been shown that, the $B_s \to \mu^+\mu^-$ and $\bar B \to \bar
K^* \mu^+ \mu^-$ decays can provide complementary information about
the anomalous $tcZ$ coupling, and therefore they are correlated with
the rare $t \to c Z$ decays. With improved measurements from the LHCb
and the future super-B factories, these interplays will be enhanced
and complementary to the direct search for the FCNC transitions in top
quark decays performed at the LHC CMS and ATLAS experiments.

\section*{Acknowledgements}
The work was supported by the National Natural Science
Foundation under contract Nos.11075059, 11225523 and 11221504. X.B.Yuan was also
supported by CCNU-QLPL Innovation Fund (QLPL2011P01).

\begin{appendix}

\section{The form factors}
For the $\bar B\to \bar K^*$ transitions, we can use seven QCD form factors
parameterize the matrix elements of $\mathcal
O_{7,9,10}$ as ($q^\mu=p^\mu-k^\mu$) \cite{FF}
\begin{align}
    \langle\bar K^*(k,\epsilon)|\bar s \gamma_\mu(1-\gamma_5)b|\bar
    B(p)\rangle=&-i\epsilon_\mu^*(m_B+m_{K^*})A_1(q^2)+i(2p-q)_\mu(\epsilon^*\cdot
    q)\frac{A_2(q^2)}{m_B+m_{K^*}}\nonumber\\
    +iq_\mu(\epsilon^*\cdot
    q)\frac{2m_{K*}}{q^2}&[A_3(q^2)-A_0(q^2)]+\epsilon_{\mu\nu\rho\sigma}\epsilon^{*\nu}p^\rho
    k^\sigma\frac{2V(q^2)}{m_B+m_{K^*}},\nonumber\\
    \langle\bar K^*(k,\epsilon)|\bar s \sigma_{\mu\nu}q^\nu(1+\gamma_5)b|\bar
    B(p)\rangle&=+T_3(q^2)(\epsilon^*\cdot
    q)\left[q_\mu-\frac{q^2}{m_B^2-m_{K^*}^2}(2p-q)_\mu\right]\nonumber\\
    +i\epsilon_{\mu\nu\rho\sigma}\epsilon^{*\nu}p^\rho
    k^\sigma 2T_1(&q^2)+T_2(q^2)\left[\epsilon_\mu^*(m_B^2-m_{K^*}^2)-(\epsilon^*\cdot q)(2p-q)_\mu\right],
\end{align}
with
\begin{align}
  A_3(q^2)=\frac{m_B+m_{K^*}}{2m_{K^*}}A_1(q^2)-\frac{m_B-m_{K^*}}{2m_{K^*}}A_2(q^2)
\end{align}
and
\begin{align}
  A_0(0)=A_3(0), \qquad\qquad T_1(0)=T_2(0).
\end{align}
At large recoil, after adopting the QCDF approach \cite{FF:lowrecoil:1}, these seven form
factors can be reduced to two universal form factors, which are related
to the form factors $V,A_{1,2}$ as \cite{FF:lowrecoil:2,QCDF:2}
\begin{align}
  \xi_\perp(q^2)=\frac{m_B}{m_B+m_{K^*}}V(q^2),  \qquad
  \xi_\parallel(q^2)=\frac{m_B+m_{K^*}}{2E_{K^*}}A_1(q^2)-\frac{m_B-m_{K^*}}{m_B}A_2(q^2).
\end{align}
At low recoil, the improved Isgur-Wise relations imply the
vector and tensor form factors are connected as follows to leading
order in $1/m_b$ \cite{Grinstein:2002,Grinstein:2004},
\begin{align}\label{eq:Isgur-Wise}
  T_1(q^2)=\kappa V(q^2), \qquad T_2(q^2)=\kappa A_1(q^2), \qquad
  T_3(q^2)=\kappa A_2(q^2)\frac{m_B^2}{q^2} ,
\end{align}
with
\begin{align}
  \kappa=1-\frac{2\alpha_s}{3\pi}\ln\left(\frac{\mu}{m_b}\right) ,
\end{align}
after neglecting subleading terms.

For the $B\to K^*$ form factors $V,A_{1,2}$, we adopt results of light-cone sum
rule approach \cite{FF}
\begin{align}
  V(q^2)&=\frac{r_1}{1-q^2/m_R^2}+\frac{r_2}{1-q^2/m_{\rm fit}^2},&{\rm
    with} \; r_1&=0.923,
  r_2=-0.511,\nonumber\\
  &&m_R&=5.32\,{\rm GeV}, m_{\rm fit}^2=49.40\,{\rm GeV^2},\nonumber\\
  A_1(q^2)&=\frac{r_2}{1-q^2/m_{\rm fit}^2},&{\rm with}\; r_2&=0.290,
  m_{\rm fit}^2=40.38\,{\rm GeV^2}\nonumber,\\
  A_2(q^2)&=\frac{r_1}{1-q^2/m_{\rm fit}^2}+\frac{r_2}{(1-q^2/m_{\rm
      fit})^2},&{\rm with}\; r_1&=-0.084, r_2=0.342, m_{\rm
    fit}^2=52.00\,{\rm GeV^2}.
\end{align}
Their relative uncertainties at $q^2=0$ are $\delta V(0)=\pm 11\%$,
$\delta A_1(0)= \pm 12\%$ and
$\delta A_2(0)=\pm 13\%$ after taking into account the uncertainties
induced by the Gegenbauer moment $a_{1,K^*}$. For the
$q^2>0$ region, the relative uncertainties are estimated to be the same as the
ones at $q^2=0$.

\section{The transversity amplitudes}
\label{sec:TA}
For the theoretical framework at the large and the low hadronic recoil region, we
follow closely the ref.~\cite{TA:largerecoil, TA:lowrecoil}
and recapitulate the relevant formulae in the following.
\subsubsection*{The transversity amplitudes at large recoil}
\label{sec:TA:large}
At large recoil, application of QCDF yields the
following transversity amplitudes \cite{TA:largerecoil}
\begin{align}\label{eq:TA:largerecoil}
A_\perp^{L,R}&=+\sqrt{2}Nm_B(1-\hat s)\left[\bigl (\mathcal C_9\mp
    \mathcal C_{10} \bigr)\xi_\perp+\frac{2\hat m_b}{\hat s}\mathcal T_\perp^+\right]\nonumber,\\
A_\parallel^{L,R}&=-\sqrt{2}Nm_B(1-\hat s)\left[ \bigl (\mathcal
  C_9\mp \mathcal C_{10}\bigr )\xi_\perp+\frac{2\hat m_b}{\hat s}\mathcal T_\perp^-\right]\nonumber,\\
A_0^{L,R}&=-\frac{Nm_B^2(1-\hat s)^2}{2m_{K^*}\sqrt{\hat s}
}\biggl[ \bigl (\mathcal C_9\mp \mathcal C_{10}\bigr )\xi_\parallel-2\hat m_b \mathcal T_\parallel^-\biggr],
\end{align}
where
\begin{align}
 \quad\hat
  s=\frac{q^2}{m_B^2}, \qquad \hat m_b=\frac{m_b}{m_B}, \qquad \hat m_{K^*}=\frac{m_{K^*}}{m_B},\qquad
  N=\left[\frac{G_F^2\alpha_{e}^2 |\lambda_t|^2}{3\cdot
      2^{10}\pi^5} m_B\hat s \sqrt{\hat\lambda}
    \beta_l\right]^{1/2} ,\nonumber\\
 \mathcal C_{9,10}=\frac{4\pi}{\alpha_s} C_{9,10},\qquad \hat \lambda=1+\hat m_{K^*}^4+\hat s^2-2(\hat
m_{K^*}^2+\hat s+\hat s \hat m_{K^*}^2).\quad\quad\quad
\end{align}
The functions $\mathcal T_{\perp,\parallel}^{\pm}$ have been
calculated at NLO \cite{QCDF:1,QCDF:2}, which have the following
general structure
\begin{align}
  \mathcal T_a^\pm&=\mathcal T_a^{\pm (t)}+ \hat\lambda_u\mathcal
  T_a^{(u)}, &  \mathcal T_a^{\pm(t)}&=\mathcal T_a^{\pm(t),{\rm
      LO}}+\frac{\alpha_s}{4\pi}\mathcal T_a^{\pm (t),{\rm NLO}},\nonumber\\
  \hat\lambda_u&= \frac{V_{ub}V_{us}^*}{V_{tb}V_{ts}^*},
&  \mathcal T_a^{(u)}&=\mathcal T_a^{(u),{\rm LO}}+\frac{\alpha_s}{4\pi}\mathcal
  T_a^{(u),{\rm NLO}},
\end{align}
where $a=\perp,\parallel$. The LO formulae read
\begin{align}
  \mathcal T_\perp^{\pm (t),{\rm LO}}&=+\xi_\perp\left(C_7^{\rm
      eff (0)}+\frac{\hat s}{2\hat m_b}Y^{(0)}\right) ,&   \mathcal T_\perp^{(u),{\rm LO}}&=+\xi_\perp\frac{\hat s}{2\hat
    m_b}Y^{(u)(0)}, \nonumber\\
  \mathcal T_\parallel^{-(t),{\rm LO}}&=-\xi_\parallel\left(C_7^{\rm
      eff (0)}+\frac{1}{2\hat m_b}Y^{(0)}\right) +HS,&    \mathcal T_\parallel^{(u),{\rm LO}}&=-\xi_\parallel\frac{1}{2\hat m_b}Y^{(u)(0)}+HS,
\end{align}
where spectator effects are denoted by $HS$. The functions involving the one-loop contributions of four-quark
operators are defined as
\begin{align}\label{eq:Y}
  Y(q^2)=&+h(q^2,m_c)\left(\frac{4}{3}C_1+C_2+6C_3+60C_5\right)
  -\frac{1}{2}h(q^2,m_b)\left(7C_3+\frac{4}{3}C_4+76C_5+\frac{64}{3}C_6\right)\nonumber\\
  &-\frac{1}{2}h(q^2,0)\left(C_3+\frac{4}{3}C_4+16C_5+\frac{64}{3}C_6\right)
  +\frac{4}{3}C_3+\frac{64}{9}C_5+\frac{64}{27}C_6,\nonumber\\
  Y^{(u)}(q^2)=&+\left(\frac{4}{3}C_1+C_2\right)\left[h(q^2,m_c)-h(q^2,0)\right].
\end{align}
The basic fermion loop function reads
\begin{align}
  h(q^2,m_q)=-\frac{4}{9}\left(\ln\frac{m_q^2}{\mu^2}-\frac{2}{3}-z\right)-\frac{4}{9}(2+z)\sqrt{|z-1|}\times
  \begin{cases}
    \arctan\frac{1}{\sqrt{z-1}} & z>1\\
    \ln {\frac{1+\sqrt{1-z}}{\sqrt{z}} }-\frac{i\pi}{2}&z\leq 1
  \end{cases}
\end{align}
with $z=4m_q^2/q^2$.
\subsubsection*{The transversity amplitudes at low recoil}
\label{sec:TA:low}
At low recoil, with the
improved Isgur-Wise form factor relations eq.~(\ref{eq:Isgur-Wise}), the transversity
amplitudes can be written as \cite{TA:lowrecoil}
\begin{align}\label{eq:TA:lowrecoil}
  A _\perp^{L,R}&=+i\left[\left(C_9^{\rm eff}\mp C_{10}^{\rm
        eff}\right )+\kappa\frac{2\hat m _b}{\hat s}C_7^{\rm eff}\right]
  f_\perp,\nonumber\\
 A_\parallel^{L,R}&=-i\left[\left(C_9^{\rm eff}\mp
   C_{10}^{\rm eff}\right)+\kappa\frac{2\hat m _b}{\hat s}C_7^{\rm eff}\right]
 f_\parallel,\nonumber\\
A_0^{L,R}&=-i\left[\left(C_9^{\rm eff}\mp C_{10}^{\rm eff}\right)+\kappa\frac{2\hat
    m_b}{\hat s}C_7^{\rm eff}\right] f_0,
\end{align}
with the definitions
\begin{align}
 N&=\left[\frac{G_{F}^2\alpha_{e}^2|\lambda_t|^2}{3\cdot 2^{10}\pi^5}m_B \hat s\sqrt{\hat
   \lambda} \right]^{\rm 1/2},&\quad
 f_\parallel&=Nm_B\sqrt 2 (1+\hat m_{K^*})A_1(q^2),\nonumber\\
 f_0&=Nm_B\frac{(1-\hat s-\hat m_{K^*}^2)(1+\hat
   m_{K^*})^2A_1(q^2)-\hat\lambda
  A_2(q^2)}{2\hat m_{K^*}(1+\hat m_{K^*})\sqrt{\hat s}},&\quad
 f_\perp&=Nm_B\frac{\sqrt{2\hat\lambda}}{1+\hat m_{K^*}}V(q^2).
\end{align}
Furthermore, we can define the two independent combinations of Wilson
coefficients as \cite{TA:lowrecoil}
\begin{align}\label{eq:rho}
 \rho_1\equiv \left|C_9^{\rm eff}+\kappa\frac{2\hat m_b}{\hat
     s}C_7^{\rm eff}\right|^2+\left |C_{10}^{\rm eff}\right |^2, \qquad \rho_2\equiv {\rm Re}\left\lbrace\left(C_9^{\rm
       eff}+\kappa\frac{2\hat m_b}{\hat s}C_7^{\rm
       eff}\right)C_{10}^{{\rm eff}*}\right\rbrace.
\end{align}
Then, at low recoil, the observables in eq.~(\ref{eq:observables}) turn
to be the transparent forms
\begin{align}
  \frac{{\rm d}\Gamma}{{\rm d}q^2}=2\rho_1\times(f_0^2+f_\perp^2+f_\parallel^2),\quad
  A_{\rm FB}=3\frac{\rho_2}{\rho_1}\times\frac{f_\perp
    f_\parallel}{f_0^2+f_\perp^2+f_\parallel^2},\quad
  F_{\rm L}=\frac{f_0^2}{f_0^2+f_\perp^2+f_\parallel^2}.
\end{align}

\end{appendix}

\end{document}